\documentclass[aps,pre,twocolumn,showpacs,superscriptaddress,groupedaddress]{revtex4}
\usepackage{amsmath}
\usepackage{graphicx}
\usepackage{soul}
\usepackage[usenames]{color}
\usepackage[abs]{overpic}
\usepackage[T1]{fontenc}
\usepackage{textcomp}

\begin{document}

\title{On-the-fly coarse-graining methodology for the simulation of chain formation of superparamagnetic colloids in strong magnetic fields}

\author{Jordi~S.~Andreu}
\email {jandreu@icmab.es}
\affiliation{Institut de Ci\`encia de Materials de Barcelona (ICMAB-CSIC), Campus UAB, E-08193 Bellaterra, Spain.}
\author{Carles Calero}
\author{Juan~Camacho}
\affiliation{Departament de F\'isica, Universitat Aut\`onoma de Barcelona, Campus UAB, E-08193 Bellaterra, Spain.}
\author{Jordi~Faraudo}
\affiliation{Institut de Ci\`encia de Materials de Barcelona (ICMAB-CSIC), Campus UAB, E-08193 Bellaterra, Spain.}

\date{\today}

\begin{abstract}
The aim of this work is the description of the chain formation phenomena observed in colloidal suspensions of superparamagnetic nanoparticles under high magnetic fields. We propose a new methodology based on an \emph{on-the-fly} Coarse-Grain (CG) model. Within this approach, the coarse grain objects of the simulation and their dynamic behavior are not fixed a priori at the beginning of the simulation but rather redefined \emph{on-the-fly}. The motion of the CG objects (single particles or aggregates) is described by an anisotropic diffusion model and the magnetic dipole-dipole interaction is replaced by an effective short range interaction between CG objects. The new methodology correctly reproduces previous results from detailed Langevin Dynamics simulations of dispersions of superparamagnetic colloids under strong fields whilst requiring an amount of CPU time orders of magnitude smaller. This substantial improvement in the computational requirements allows the simulation of problems in which the relevant phenomena extends to time scales inaccessible with previous simulation techniques. A relevant example is the waiting time dependence of the relaxation time $T_2$ of water protons observed in magnetic resonance experiments containing dispersions of superparamagnetic colloids, which is correctly predicted by our simulations. Future applications may include other popular real-world applications of superparamagnetic colloids such as the magnetophoretic separation processes. 
\end{abstract}


\maketitle

\section{Introduction}

In recent years, work in coarse grain models for the description of soft matter and biomolecular systems is experiencing a remarkable outburst~\cite{CGreview}. The reason is that the description of these systems at experimentally relevant time and length scales requires inclusion of phenomena occurring at very different scales. The objective of coarse grain (CG) models is thus to retain sufficient molecular or nanoscale detail and yet remain amenable of simulation up to macroscopic time scales. Many approaches have been developed to construct CG models of different kinds of soft matter systems. For example, in the case of polymers, there is a long tradition of using CG models and the field is sufficiently mature so that there are systematic and rigorous approaches to build up CG models from accurate atomistic descriptions~\cite{Pep}. Also, in the field of biomolecular simulations, there are important developments such as the MARTINI force field~\cite{Martini} which allow the simulation of difficult problems such as the behavior of lipid vesicles or protein folding at millisecond or even larger time scales. New advances include also simulation packages specially designed for CG models of soft matter such as ESPResSo~\cite{Espresso}. 

Our interest here is the development of an improved CG model for a specific problem which is still difficult to simulate, namely the assembly of superparamagnetic colloids under strong magnetic fields. Superparamagnetic colloids are made of small nanoparticles of magnetic material (typically 5-10 nm iron oxide nanocrystals) embedded in a nonmagnetic matrix (typically polymers or silica)~\cite{Taboada2009}. These particles have no magnetic dipole in absence of magnetic field but they develop very high magnetizations in the presence of a magnetic field, similar to those obtained with ferromagnetic materials. This highly tunable response and the possibility to functionalize their surface make these materials very interesting for applications such as capture and removal of biomolecules and pollutants, NMR contrast agents, and many others~\cite{Yavuz2009,Corchero2009,Krishnan2010}. 

Our work is motivated by the difficulties encountered in modeling different kinds of real experimental situations involving superparamagnetic colloids. A relevant example is provided by the experiments by Chen et al.~\cite{Chen2011} of a dispersion containing superparamagnetic colloids designed as contrast agents for magnetic resonance imaging (MRI). In these experiments, a strong uniform magnetic field was applied to the dispersion. It was observed that the transverse relaxation time $T_2$ of protons in water changed with time, an effect which was attributed to the formation of chains of superparamagnetic colloids. In fact, the kinetics of chain formation was estimated from these experiments, spanning time scales from 10 to 10$^3$ seconds or more. Another relevant example is magnetophoresis~\cite{Faraudo2010}, which is the motion of magnetic particles under an external magnetic gradient. Experimental evidence shows that the formation of chains induced by the external field speeds up the magnetophoresis process~\cite{DeLasCuevas2008}, which is orders of magnitude faster than that observed in absence of chain formation~\cite{Andreu2011b}. It is worth noting that in these experiments, chains dissolve almost immediately after removal of the external field, as should be expected since superparamagnetic nanoparticles have no dipole in absence of magnetic field. In this respect, these systems are very different from the widely studied (and simulated) dispersions of dipolar particles, which are able to form structures in the absence of an external magnetic field due to the interaction of their permanent dipoles~\cite{DeGennes1970,Helgesen1988,Ivanov2004}.

The standard approach for simulation of chaining processes in magnetic colloids is the use of Langevin Dynamics (LD) simulations (see for example~\cite{Dominguez2007,Andreu2011}). This technique allows the inclusion of particle-particle interactions, thermal noise and the friction due to the fluid. The resolution of the simulation technique is typically in the nanoseconds scale. Simulation runs up to a few seconds are possible, but they are highly intensive, requiring the use of parallel computing during several weeks~\cite{Andreu2011}. These CPU requirements make this simulation technique unsuitable for the study of the problems mentioned above.

The need to account for microscopic time and length scales but also reach macroscopic time scales at low computational cost has motivated us the development of a new simulation strategy based on an \emph{on-the-fly} CG procedure. The methodology which will be developed in this paper is a generalization of two procedures proposed in previous works: the method proposed by Miguel and Satorras~\cite{Miguel1999} to study aggregation processes and the method proposed by Schaller et al.~\cite{Schaller2008} to study magnetophoresis.   

In the methodology proposed here, one starts by simulating the motion and interaction between individual colloids. As the simulation advances, colloids form chains due to the magnetic dipole-dipole attraction induced by the high external magnetic field. The motion of each particle inside a chain is not simulated explicitly. In our methodology, these chains are considered individual coarse grain (CG) objects which move following certain effective rules and interact (and possibly aggregate) with other CG objects or single individual particles. In this way, the CG objects  of the simulation are not fixed {\it a priori} at the beginning of the simulation but rather redefined \emph{on-the-fly}. Thus, we adjust the resolution of the calculations during the simulation run, allowing for the possibility of much longer simulation runs requiring less computer power. Preliminary simulation results and comparison with experiments, presented in a previous work~\cite{Chen2011}, demonstrated the feasibility and utility of our novel approach. Here we will discuss in detail the physical basis of the model, the simulation methodology and detailed comparison with more standard Langevin simulation techniques. All simulations of our model were performed employing the MagChain program, a C++ application developed in-house, which is freely available for use of researchers. The code, its documentation and usage examples can be found available for download at our web page~\cite{Web}.

The paper is organized as follows. In Section II, we describe the modeling of the system under study and the simulation technique. In Section III we validate the new methodology by (i) comparing our results with those obtained employing standard LD simulations (ii) discussing the effect of choosing other approximations for the diffusion coefficients and the effective interaction of the CG objects and (iii) discussing the applicability of our methodology comparing with experimental results. The conclusions are presented in section IV and some technical issues are detailed in the Appendix.

\section{Formulation of the Model}

The system which we are interested to describe is a colloidal dispersion of $N_p$ superparamagnetic spherical particles of diameter $d$ in a volume $V$ and volume fraction $\phi_0$:
\begin{equation}
\phi_0=\frac{N_p}{V} \frac{\pi}{6} d^3.\label{phi}
\end{equation}
In absence of external magnetic field, the particles have no magnetic dipole and there is no formation of chains (no aggregation induced by the magnetic field). In the presence of a magnetic field $H$, the superparamagnetic particles acquire a certain magnetization $M(H)$. Since we are particularly interested in the case of very strong magnetic fields (as in the experiments of Ref.~\cite{Chen2011} for example), we consider that the particles have a magnetic dipole moment $m_s$ (corresponding to the saturation magnetization $M_s$ of the particles) pointing in the direction of the applied magnetic field (which we will take as the $z$ axis). The strength of the magnetic interaction between particles as compared with thermal energy can be characterized by the magnetic coupling parameter $\Gamma$ defined as:
\begin{equation}
\Gamma=\frac{\mu_0m_s^2}{2\pi d^3k_BT}.\label{gamma}
\end{equation}
The behavior of superparamagnetic colloids under external fields is controlled by the values of these two parameters ($\phi_0$ and $\Gamma$). In this paper, we are interested in situations (values of $\phi_0$ and $\Gamma$) in which the external field induces formation of linear chains of colloids which grow irreversibly with time. Irreversible growth of linear chains has been found in simulations and experiments investigating the ranges of $\Gamma$ between 40 and 3$\times 10^3$ and $\phi_0<0.15$ \cite{Fermigier1992,Dominguez2007,Andreu2011}. However, other structures are found at different ranges of  $\phi_0$ and $\Gamma$. For lower values of $\Gamma$, an equilibrium state is possible, in which colloids aggregate in linear (non branched) chains with an equilibrium length given by $\sqrt{\phi_0 e^{\Gamma-1}}$ \cite{Andreu2011}. In the opposite situation of larger values of $\phi_0$ and $\Gamma$, different aggregate structures can be found, including thick chains (obtained from lateral aggregation of linear chains),  bundles and more complex fibrous structures \cite{Fermigier1992}. All these more complex situations, different from irreversible growth of linear chains, will be left for future extensions of the model.

As key ingredients to retain the underlying physics of irreversible chain growth, we consider both the diffusive motion of particles and chains and their respective magnetic and steric interactions. The main approximations of our model will be to ignore the details of the particles forming the chains and to replace the actual magnetic dipole-dipole interaction by an effective short-range interaction, less demanding from the computational point of view.

\begin{figure*}[htp]
\begin{center}
\includegraphics*[width=18cm]{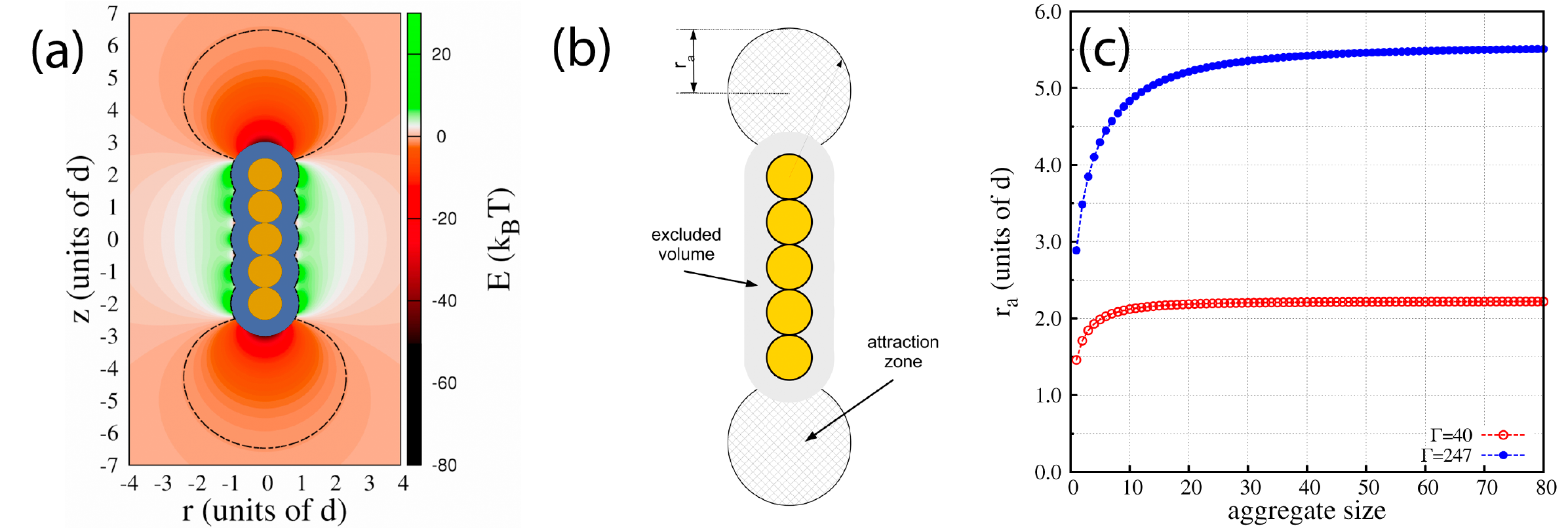}
\caption{(Color online) \textbf{(a)} 2D-map corresponding to the interaction energy between an incoming test dipolar particle and a chain-like aggregate formed by 5 colloids with a magnetic coupling of $\Gamma$=40. The black dashed line delimites the region with ($E<-k_BT)$. The region excluded by the finite size of the 5 spheres is shown in blue (the interaction energy map is not evaluated inside this region). \textbf{(b)} Sketch (top view) of the attraction model implemented in the MagChain code. Each CG object has two attraction zones, modeled as a sphere  of radius $r_a$ (Eq.~(\ref{ra}) tangent to the edge of the aggregate. Any particle entering into these zones will immediately aggregate forming a longer chain. \textbf{(c)} Dependence of the radius ($r_a$) of the attraction regions on the aggregate size for two different values of the coupling parameter, $\Gamma$=40 (open symbols) and $\Gamma$=247 (solid symbols). The attraction radius increases abruptly for short chains and tends to a constant value for longer chains. All the distances are expressed in terms of the diameter $d$ of the colloid.}
\label{fig:map}
\end{center}
\end{figure*}
Our model to study the kinetics of chain formation in these systems consists of CG objects which are chains made of $s$ particles, including the case $s=1$ which corresponds to a single particle. The first ingredient of the model is the diffusion coefficient of the CG objects. For single, isolated particles ($s=1$) we have:
\begin{eqnarray}
D_1 = \frac{k_BT}{3\pi\eta d} \label{DReal},
\end{eqnarray}
where $\eta$ is the viscosity of the fluid. A chain containing $s>1$ particles exhibits anisotropic diffusion, characterized by a diffusion coefficient $D_s^{\parallel}$ in the direction parallel to the long axis of the chain and $D_s^{\perp}$ in the directions perpendicular to the long axis. There are several possibilities for the analytical form of these diffusion coefficients, depending on the exact geometry assumed for the chains and the degree of approximation of the calculation. Here, in order to keep the model as simple as possible, we consider the following expressions valid for elongated objects (slender body theory~\cite{Brenner}):
\begin{equation}
\label{DparER}
\frac{D_s^{\parallel}}{D_1} = \frac{3}{2s}[\ln(2s) - \frac{1}{2}],
\end{equation}
\begin{equation}
\label{DperpER}
\frac{D_s^{\perp}}{D_1} = \frac{3}{4s}[\ln(2s) + \frac{1}{2}].
\end{equation}
Strictly speaking, Eqs.~(\ref{DparER}) and (\ref{DperpER}) are valid only for large $s$. Therefore, we employ Eqs.~(\ref{DparER}) and (\ref{DperpER}) for chains with $s>2$ and use a simple interpolation between the diffusion coefficients corresponding to CG objects with $s=1$ (Eq.~\ref{DReal}) and $s=3$ (Eqs.~(\ref{DparER}) and (\ref{DperpER})) for chains with $s=2$. Such a choice gives results indistinguishable from those provided by more sophisticated and accurate expressions of the diffusion coefficients (see Section IIIB).

The second ingredient of the model is the definition of the effective interaction between CG objects. A CG object of length $s$ interacts with other CG objects through an excluded volume interaction (hard core) corresponding to a cylinder of length $s\times d$ and diameter $d$. They also interact through dipole-dipole interactions. In order to simplify and speed up the simulations, we have replaced the actual dipole-dipole magnetic interaction between colloids by an effective, short range interaction between the CG objects. This interaction is defined as follows. For a given CG object, we define two spherical attractive regions of radius $r_a(s)$ (which depend on the length of the chain $s$) located at the two ends of the chain. As illustrated in Fig.~\ref{fig:map}, these regions are designed to mimic the region at which the magnetic attraction between a chain of particles (magnetized in the $z$ direction) and an incoming test dipolar particle is equal or stronger than the thermal energy $k_BT$. The values of $r_a(s)$ are calculated by finding the distance in the $z$ axis at which the magnetic interaction energy $E_{mag}$ between a chain of $s$ particles and a single test particle is equal to $-k_BT$. Therefore, $r_a(s)$ is given by the solution of:
\begin{equation}
\label{ra}
 \frac{E_{mag}}{k_BT}=-\Gamma \sum_{n=0}^{s-1} \frac{1}{(2r_a(s)/d+1/2+n)^3}=-1.
\end{equation}
The results of Eq.~(\ref{ra}) for different values of $\Gamma$ are also shown in Fig.~\ref{fig:map}. Once we have defined the range of the interaction, we need to define the strength of this interaction. In order to keep our model as simple as possible, we simply assume that all events in which a CG object enters into the interaction region of another CG object will lead to instantaneous aggregation. This rule has been employed previously in the interpretation of experimental results and it has been suggested by direct observations of chain formation under a microscope (see Refs.~\cite{Promislow1995,Martinez-Pedrero2007}). As we will show in the next section, this rule reproduces correctly the results obtained from LD simulations in which the magnetic interaction is computed accurately. The sensitivity of the results to the choice of $r_a$ will be also discussed in Section IIIB.

\begin{figure}[htp!]
\begin{center}\includegraphics*[width=9cm]{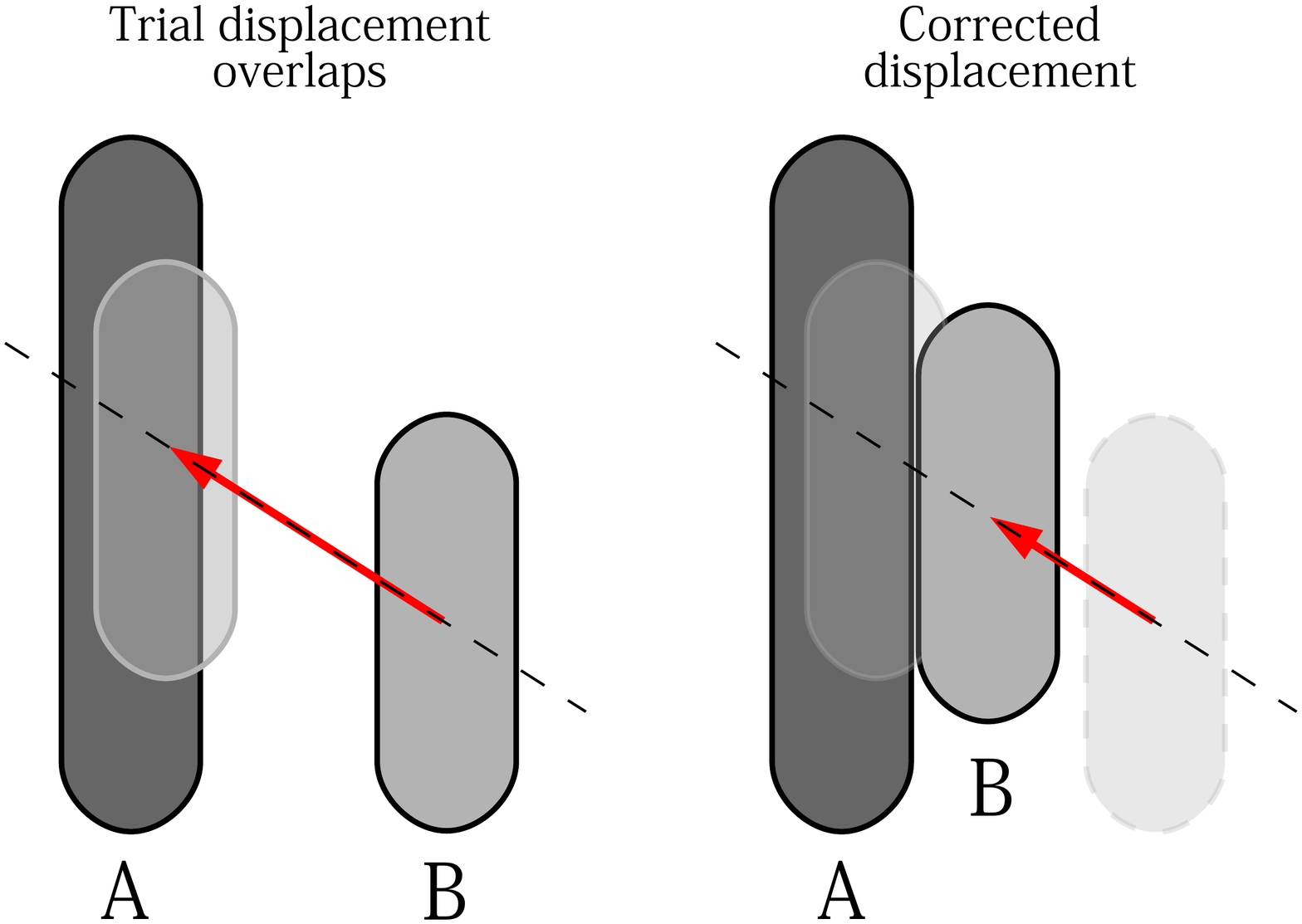}
\caption{(Color online) Sketch corresponding to the scheme applied to avoid the the overlap between CG objects during simulations. If the random displacement performed on object B produces an overlap between B and another CG object (A), the moving aggregate (B) is placed in contact with the second aggregate (A) according to the trajectory followed during the random displacement.}
\label{fig:collision}
\end{center}
\end{figure}

Once the basic ingredients for the model (rules for motion and interaction) are defined, it is necessary to specify the algorithm for the numerical solution of the model. In our case, the diffusive motion of the CG objects is simulated using the brownian dynamics technique~\cite{BD}.  At each time step $\Delta t$ a random displacement in each direction is generated with a gaussian distribution with zero mean and variance 2$D_s \Delta t$, where $D_s$ is the diffusion coefficient of the object (single particle or chain) in the direction of motion ($x$, $y$ or $z$). Also, at each time step the distances between CG objects are checked in order to detect penetration of a CG object inside the region of aggregation of another CG object, as explained above, or to detect possible overlaps between them. In the case of aggregation of two CG objects, a new CG object is created (and the two previous CG objects are erased from the simulation) with length $s$, obtained from adding the lengths of the two aggregating chains and located at the center of mass of the aggregating CG objects. In the case of overlap between two CG objects without penetration into the aggregation region, we consider that the two chains collide. In this case, the moving CG object is placed in contact with the other one (without overlapping) at the collision coordinates defined by the trajectory previously followed (see Fig.~\ref{fig:collision}). Finally, it should be noted here that the selection of an appropriate time step $\Delta t$ for the simulation is a crucial issue. A detailed discussion on the selection of $\Delta t$ is given in the Appendix.

Hence, a typical simulation run is as follows. The simulation starts from a pre-equilibrated system containing $N_p$ colloids (CG objects with $s=1$). As the simulation goes on, colloids aggregate and chains with increasing values of $s$ appear. Consequently, the number of CG objects of the simulation decreases with time and the simulation speeds up as the time advances, as we will discuss in detail in the following section. As a simulation output, we obtain the number of chains containing $s$ colloidal particles at time $t$, $n_s(t)$. During the simulation, we also monitor the time evolution of the average number of colloidal particles in a chain $\langle N\rangle(t)$ defined as in~\cite{Miguel1999,Dominguez2007}:
\begin{eqnarray}
\label{N}
\langle N(t)\rangle=\frac{\sum_s s n_s(t)}{\sum_s n_s(t)} = \frac{N_p}{\sum_sn_s(t)},
\end{eqnarray}
and the probability of finding an aggregate of size $s$ at a given time, defined as:
\begin{eqnarray}
\label{P}
p(s; t) = \frac{n_s(t)}{\sum_sn_s(t)}.
\end{eqnarray}

\section{Validation and application of the model}
\subsection{Comparison with Langevin Dynamics simulations}

\begin{table*}[htdp]
\caption{Characteristics of the colloidal dispersions of superparamagnetic particles simulated with CG and LD techniques. $\phi_0$ and $\Gamma$ are defined by Eqs.~(\ref{phi}) and (\ref{gamma}), $\rho_p$ is the density of a single colloid, $d$ is its diameter and $D_1$ its diffusion coefficient. $T$ is the temperature of the dispersion and $\eta$ the viscosity of the solvent (water) at this temperature.}
\begin{center}
\begin{tabular}{|c|c|c|c|c|c|c|c|c|c|}
\hline
 & $\Gamma$ & $\phi_0$ & $\rho_p$ [g/cm$^3$] & $d$ [nm] & $T$ [K] & $\eta$ [$Pa\cdot s$] & $D_1$ [m$^2$/s]\\
\hline
Case 1 & 40 & 5.23$\times 10^{-4}$& 1.0 & 100 & 300 & 1.0$\times 10^{-3}$ & 4.39 $\times 10^{-12}$ \\
Case 2 & 247 & 4.64$\times 10^{-6}$& 3.1 & 88 & 310 & 0.692$\times 10^{-3}$ & $7.46 \times 10^{-12}$ \\
\hline
\end{tabular}
\end{center}
\label{tab:colloids}
\end{table*}
\begin{table*}[htdp]
\caption{Set of parameters used for numerical integration in the Coarse Grain (CG40 and CG247) and the Langevin Dynamics (LD40 and LD247) simulations. $N_p$ is the number of particles in the simulation, $L_z$ and $L_x=L_y$ are the sizes of the simulation box (in units of particle diameters) in the directions parallel and perpendicular to the magnetic field, respectively (periodic boundary conditions were employed in all simulations). $\Delta t$ is the time step and $t_f$ is the total simulated time. We also indicate the total amount of CPU time employed in the calculation, calculated as the number of cores used times the total elapsed time. In all our calculations we have used a 8-Core AMD Opteron Magny Cours 6136 processor.}
\begin{center}
\begin{tabular}{|c|c|c|c|c|c|c|c|c|c|c|c|}
\hline
Label & System &$N_p$ & $L_z$ & $L_x=L_y$ & $\Delta t$ [s] & $t_f$ [s] & \# cores & CPU cost \\
\hline
LD40 & Case 1& 8000 & 200 & 200 & 1.02$\times 10^{-9}$ & 2.04 & 8 & 998h \\
LD247 & Case 2& 4000 & 767 & 767 & 3.06$\times 10^{-9}$ & 6.12 & 8& 866h \\
CG40 & Case 1& 8000 & 512 & 128 & $2.280\times 10^{-4}$ & 5 & 1 & 25h \\
CG247 & Case 2& 8000 & 1534 & 767 &  $1.038\times 10^{-4}$& 1000 & 1 & 24h \\
\hline
\end{tabular}
\end{center}
\label{tab:parameters}
\end{table*}

\begin{figure*}[htp!]
\begin{center}\includegraphics*[width=18cm]{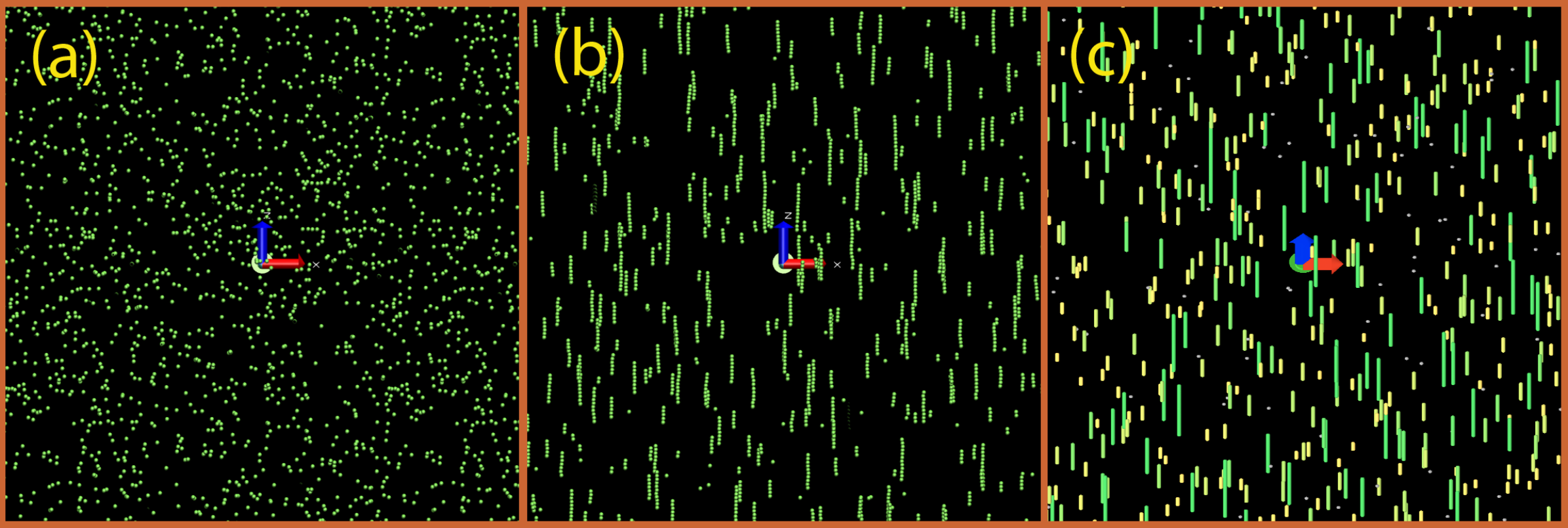}
\caption{(Color online) Snapshots from simulations with $\Gamma=40$ (Case 1 of Table~\ref{tab:colloids}). \textbf{(a)} Initial configuration of a simulation ($t=0$) \textbf{(b)} Snapshot from Langevin Dynamics simulations (LD40) at $t=0.28$s. Note that the simulation resolves the individual particles building up the chains. \textbf{(c)} Snapshot from Coarse Grain simulations at $t=0.28$s. Now the chains are the CG objects, individual particles are no longer considered once they form part of a chain. Chains are colored according to their length for an easier visualization. Left and center images were created using VMD~\cite{VMD}. Right image was created using our own visualization software available in the web~\cite{Web}.}
\label{fig:snapshot}
\end{center}
\end{figure*}

\begin{figure}[htp]
\begin{center}
\includegraphics*[width=8cm]{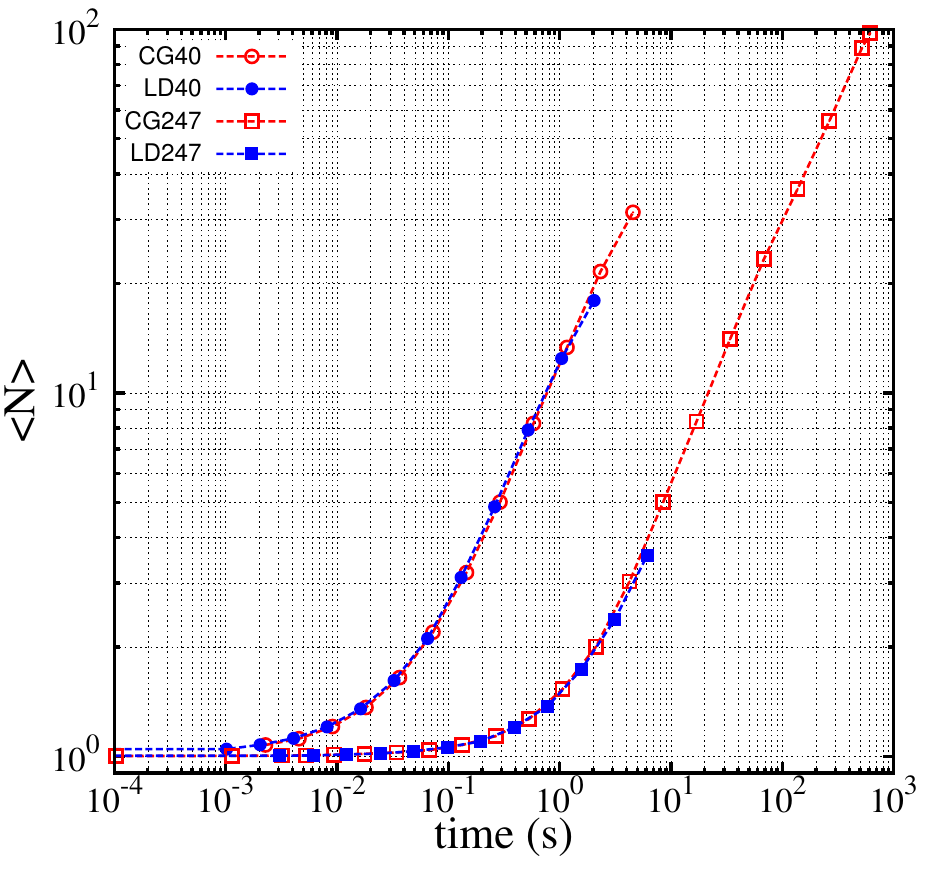}
\caption{(Color online) Time evolution of the average number of particles $\langle N(t)\rangle$ in a chain, Eq.~(\ref{N}). Comparison between the results obtained from Langevin Dynamics (solid symbols) and Coarse Grain (open symbols) simulations for the two different systems studied. Circles correspond to Case 1 ($\Gamma$=40,$\phi_0$=5.23$\times 10^{-4}$) and squares to Case 2 ($\Gamma$=247, $\phi_0$=4.64$\times 10^{-6}$).}
\label{fig:N}
\end{center}
\end{figure}

\begin{figure}[htp]
\begin{center}
\includegraphics*[width=9cm]{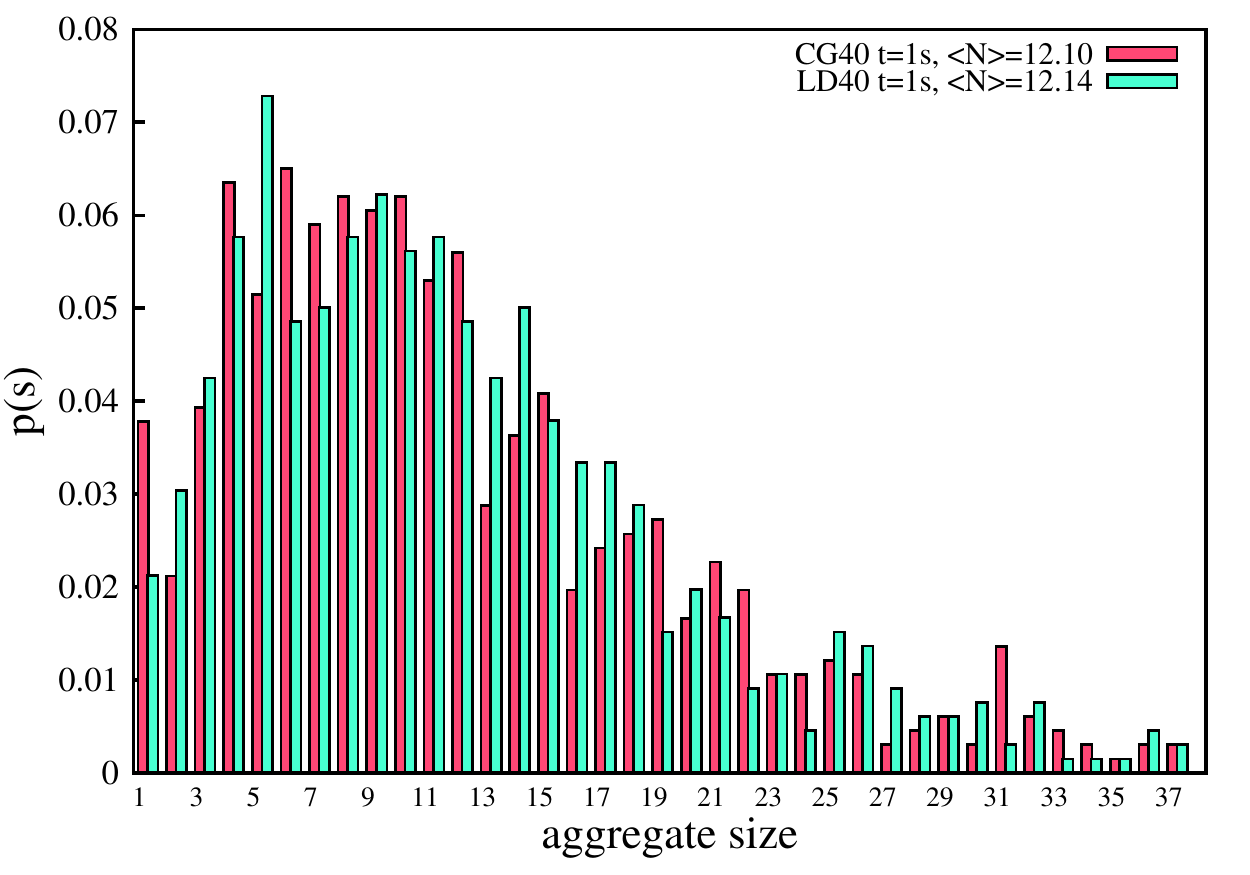}
\caption{(Color online) Comparison between the probability distribution to find an aggregate of size $s$ (defined in Eq.~(\ref{P})) at t=1s obtained from LD40 and CG40 simulations.}
\label{fig:comp_dist_G40}
\end{center}
\end{figure}

\begin{figure}[htp]
\begin{center}
\includegraphics*[width=9cm]{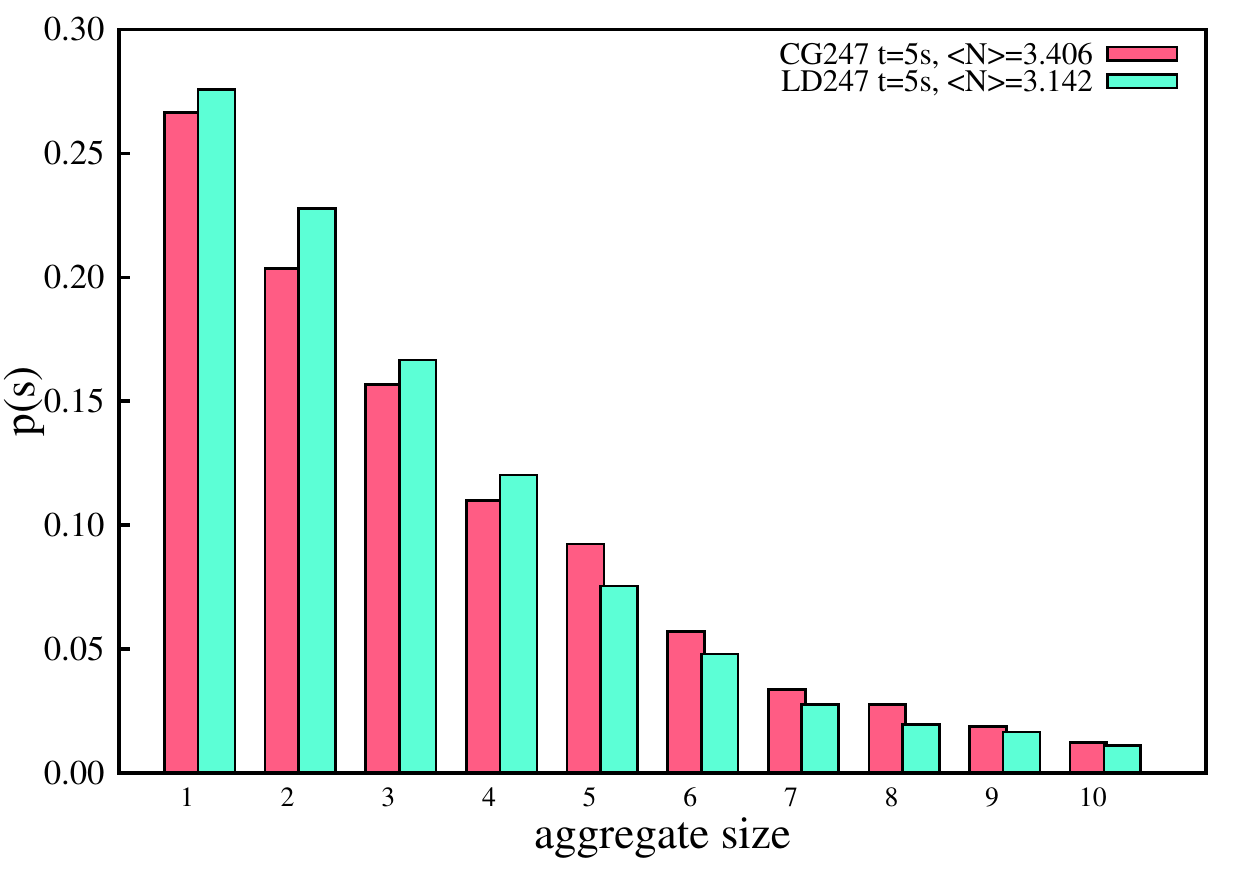}
\caption{(Color online) Comparison between the probability distribution to find an aggregate of size $s$ at t=5s  obtained from LD247 and from CG247.}
\label{fig:comp_dist_G247}
\end{center}
\end{figure}

Our objective in this section is to compare the performance and results obtained using the model described in Section II with standard LD simulations of the same system. Briefly stated, Langevin Dynamics simulations consist on solving the Newton equations of motion for each particle taking into account external forces, the interaction forces between particles (magnetic and steric), the viscous drag from the solvent and a stochastic force arising from the thermal noise due to the fact that the system is at a given temperature $T$. This comparison between our simplified CG method and more detailed LD simulations will help to clarify the validity of the approximations introduced in our model, as described in the previous section. In order to perform a significant comparison between the new procedure and the standard LD simulation technique, we have selected two cases with very different magnetic coupling $\Gamma$ which were studied in previous works. The details for these systems are summarized in Table~\ref{tab:colloids} and were denoted as Case 1 and Case 2.

Let us consider first Case 1, which corresponds to a dispersion of 100 nm superparamagnetic colloids at a volume fraction $\phi_0=5.23\times10^{-4}$ which have a magnetic coupling parameter $\Gamma=40$ at saturation (i.e. under strong magnetic fields). This system was studied employing LD simulations in Ref.~\cite{Andreu2011} by using the standard LAMMPS simulation package~\cite{LAMMPS} (version 21May2008). Now, we will compare these published results obtained with the standard LD technique with our CG methodology described in the previous section. These simulations will be denoted as LD40 (the Langevin Dynamics case) and CG40 (our Coarse Grain methodology). The parameters employed in the numerical algorithm are given in Table~\ref{tab:parameters}. For completeness, we also give the parameters employed in the LD simulations. It should be noted that the LD simulations require a small time step (of the order of the ns). This small time step is needed in order to avoid a simulation crash in simulations involving chains of colloids, since the motion of colloids inside a chain involves very small displacements which need to be resolved with high precision. In our CG methodology (which do not consider the structure of chains), we can use much larger time steps as shown in Table~\ref{tab:parameters}. A detailed discussion on the selection of the time step in our methodology is given in the Appendix.

Typical snapshots of the simulation are shown in Fig.~\ref{fig:snapshot} and the results for $\langle N(t)\rangle$ are shown in Fig.~\ref{fig:N}. The snapshots illustrate the different resolution  employed in the CG40 and LD40 simulations. As seen in the snapshots, the LD40 simulation resolves the individual particles making up the chains whereas the chains are structureless in the CG40 simulation. It should be noted that the chains obtained in the LD40 simulation are almost perfectly linear and are not significantly different from the coarse-grain objects of the CG40 simulation. As shown in Fig.\ref{fig:N}, the values of $\langle N(t)\rangle$ obtained from both simulations (CG40 and LD40) are in excellent agreement. For example, at $t=1$s, the mean aggregate size for the CG40 simulation is $\langle N\rangle_{CG}$=12.10 and the value calculated from LD simulations is almost identical, $\langle N\rangle_{LD}$=12.14. Therefore, we can conclude that the simplifying approximations included in the new methodology (particularly those regarding the calculations of particle-particle magnetic interaction) do not affect the average size of chains. 

We have performed a more detailed comparison between both approaches by comparing the distribution of chains of size $s$ at certain times. 
In Fig.~\ref{fig:comp_dist_G40}  we compare the corresponding probability distribution (defined in Eq.~(\ref{P})) at $t=$1s obtained from LD40 and CG40 simulations. The agreement between both results after 1s is remarkable, and only slight differences are observed. As shown in Fig.~\ref{fig:comp_dist_G40}, the distribution of chain sizes is very broad, with significant probabilities of finding chains well above and well below the average length (including isolated particles).

Now, let us consider the case denoted as Case 2 in Table~\ref{tab:colloids} corresponding to one of the samples considered in the experiments in~\cite{Chen2011}. In this case, the particles have larger saturation magnetization ($\Gamma=247$) but the dispersion is more diluted ($\phi_0=4.64\times10^{-6}$). We have performed Langevin Dynamics simulations as well as simulations employing our new methodology, as in Case 1. These simulations will be denoted as LD247 and CG247, respectively. A list of relevant simulation parameters for LD247 and CG247 simulations is given in Table~\ref{tab:parameters}. The results obtained for $\langle N(t)\rangle$ are also given in Fig.~\ref{fig:N}. The results for the probability distribution $p(s;t)$ at $t=5$s are shown in Fig.~\ref{fig:comp_dist_G247}. Again, we obtain a good agreement between the predictions of both simulation methodologies, both in the average size of chains and in the probability distribution of chain sizes.

As shown in Table~\ref{tab:parameters}, both for the case with $\Gamma=40$ and $\Gamma=247$, the computational cost of the CG simulation technique is much lower than the corresponding LD simulation. For example, we note that a production run of 6 seconds for the LD247 simulation requires about 4.5 days of calculations, with the program running in parallel in a 8-Core AMD Opteron Magny Cours 6136 processor. In contrast, the CG247 simulation requires less than 4 hours to simulate the same physical time using a single core of the same processor employed in the LD247 run. In addition, we can reach surprisingly long time scales in our CG247 simulation with a very low computational cost (see Table~\ref{tab:parameters}). In simulation CG247, we reach simulated times up to $10^3$ s in a 1 day calculation, a time scale two orders of magnitude larger than that accessible using Langevin Dynamics simulations.

\begin{figure}[htp]
\begin{center}
 \begin{overpic}[width=9cm]{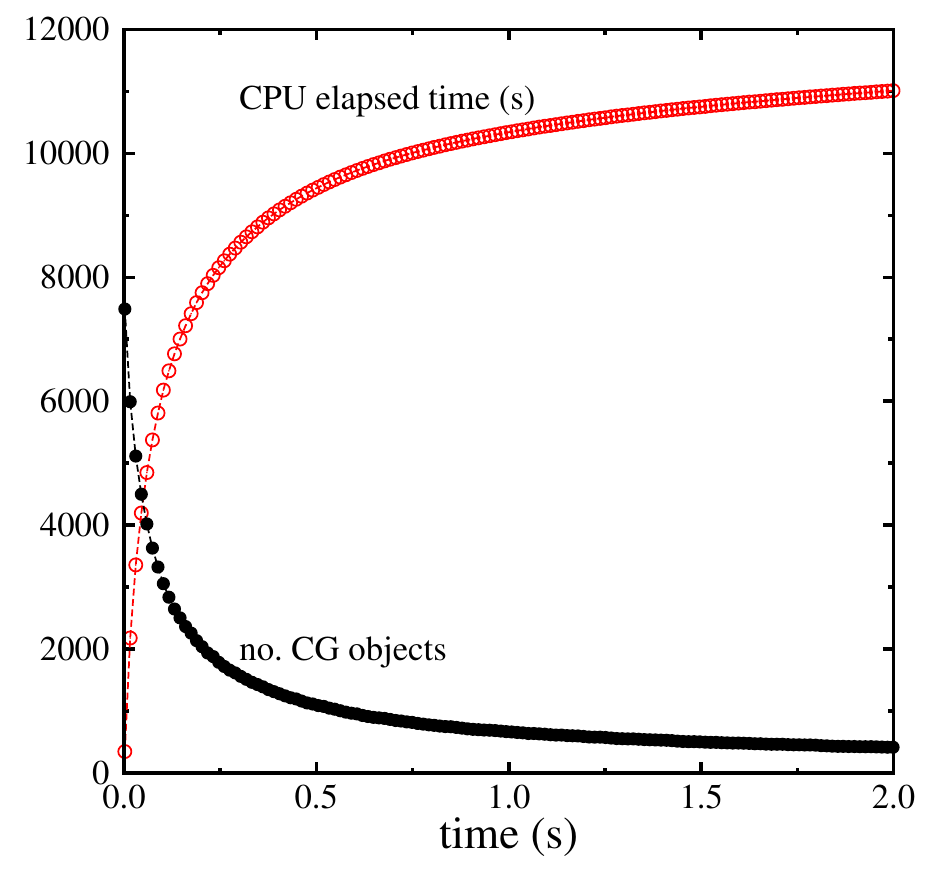}
     \put(97,58){\includegraphics[width=5.2cm]{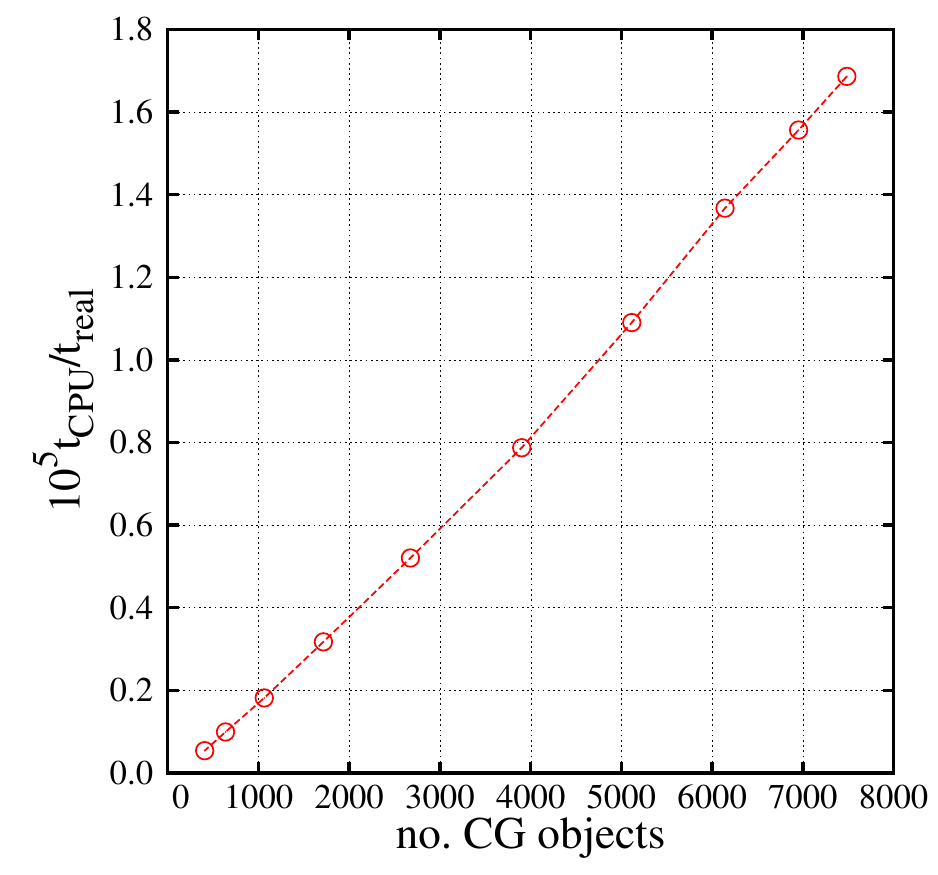}}
    \end{overpic}
\caption{(Color online) Elapsed CPU time (open circles) and number of CG objects (solid circles) as a function of the real simulated time for the CG40 simulation. \textbf{Inset: }Rate between the elapsed CPU time and the real simulated time as a function of the number of CG objects present in the CG40 simulation.}
\label{fig:performance_CG40}
\end{center}
\end{figure}
The CPU costs shown in Table~\ref{tab:parameters} demonstrate that the new simulation technique allows us to perform simulations of the two systems considered here with an extremely reduced computational effort as compared to Langevin Dynamics simulations. Moreover, it is also important to notice that the required CPU time for the CG approach to simulate a certain time interval is reduced during a simulation, since the number of CG objects decreases as the simulation advances. This effect is clearly shown in Fig.~\ref{fig:performance_CG40} for the CG40 simulation. We also show that the rate between the elapsed CPU time and the corresponding real time simulated depends linearly with the number of CG objects (see inset Fig.~\ref{fig:performance_CG40}). It should be noted that in the LD simulations the opposite effect was observed; the fact that the individual motion of the particles inside the chains are fully resolved makes LD simulations increasingly inefficient as time goes on.

\subsection{Further discussion on the approximations of the model}

\subsubsection{Diffusion Model}

\begin{figure}[htp]
\begin{center}
\includegraphics*[width=8cm]{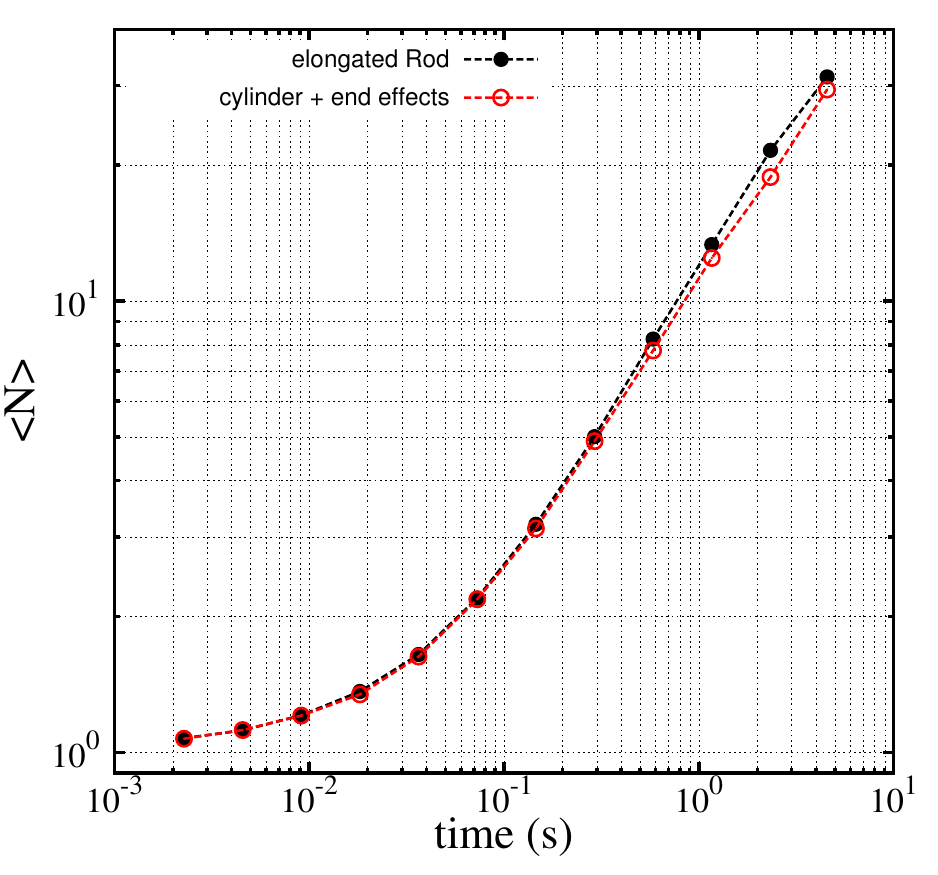}
\caption{(Color online) Comparison between the results obtained from the elongated rod approximation (corresponding to Eqs.~(\ref{DparER}) and (\ref{DperpER}) and represented by solid circles) and the cylinder approximation with end-effects (corresponding to Eqs.~(\ref{DparCylEE}) and (\ref{DperpCylEE}) and represented by open circles) when computing the mean chain size of the aggregate $\langle N\rangle$ for the CG40 system.}
\label{fig:comp_diffusion}
\end{center}
\end{figure}

As it has been already mentioned, one of the two key ingredients of the CG approach is the diffusion model adopted to describe the motion of the coarse grain objects. In order to check the possible influence of the model selected, we have also performed simulations with a different diffusion model proposed by Tirado et al.~\cite{Tirado1979} in which they describe the translational motion of right circular cylinders also accounting for the so called \textit{end-effects}. 
Following the same approach than in~\cite{Martinez-Pedrero2007}, we have used the expressions:
\begin{eqnarray}
\frac{D_s^{\parallel}}{D_1} = \frac{3}{2s}[\ln(s) + \gamma^{\parallel}(s)],\label{DparCylEE}\\
\frac{D_s^{\perp}}{D_1} = \frac{3}{4s}[\ln(s) + \gamma^{\perp}(s)].\label{DperpCylEE}
\end{eqnarray}
where $\gamma^{\parallel}$ and $\gamma^{\perp} $are the \textit{end-effect} functions defined as:
\begin{eqnarray}
\gamma^{\parallel}(s) = - 0.21 + \frac{0.90}{s},\label{ee-parallel}\\
\gamma^{\perp}(s) = 0.84 + \frac{0.18}{s} + \frac{0.24}{s^2}.\label{ee-perp}
\end{eqnarray}

We have computed the mean aggregate size $\langle N\rangle$ using both diffusion models for the CG40 system and the results obtained are plotted in Fig.~\ref{fig:comp_diffusion}. As it can be seen from these results, no significant differences are found in the average number of particles $\langle N\rangle$ obtained with both diffusion models. For this reason we can conclude that both models are suitable for the description of the diffusive motion of the chain-like aggregates in such systems and our selection of the elongated rod model (Eqs.~(\ref{DparER}) and (\ref{DperpER})) instead of Eqs.(\ref{DparCylEE})-(\ref{ee-perp}) for the simulations was based on its major simplicity.

\subsubsection{Effective interaction. Attraction radius}
As explained in detail in Section II, we have defined the aggregation regions for each CG object as the surrounding space in which the magnetic interaction energy between the CG object and a dummy single particle is equal or smaller than $-k_BT$. As shown in Fig.~\ref{fig:map}, the attraction radius $r_a$ defining this region depends on the size of the considered aggregate and on the magnetic coupling parameter $\Gamma$ (see Eq.~\ref{ra}). It is observed that for small chains the attraction radius increases with their size and tends to a constant value for larger chains (the addition of a new particle into the same aggregate does not significantly contribute to the interaction magnetic energy). Here, we would like to demonstrate the importance of accounting for the $s$ dependence of $r_a(s)$ in the simulations.

To this end, we have performed two additional simulations in which the  $s$ dependence of $r_a(s)$ is ignored. A first simulation (denoted as CG40-min) corresponds to a repetition of the simulation CG40 of the previous section (see Table~\ref{tab:parameters}) but using $r_a=1.46d$ for all chains. We also performed another simulation (denoted as CG40-max) in which we employed the value $r_a=2.20d$ for all chains. These values correspond to the minimum and maximum values of $r_a(s)$ employed in the original CG40 simulation (see Fig.~\ref{fig:map}).

\begin{figure}[htp]
\begin{center}
\includegraphics*[width=8cm]{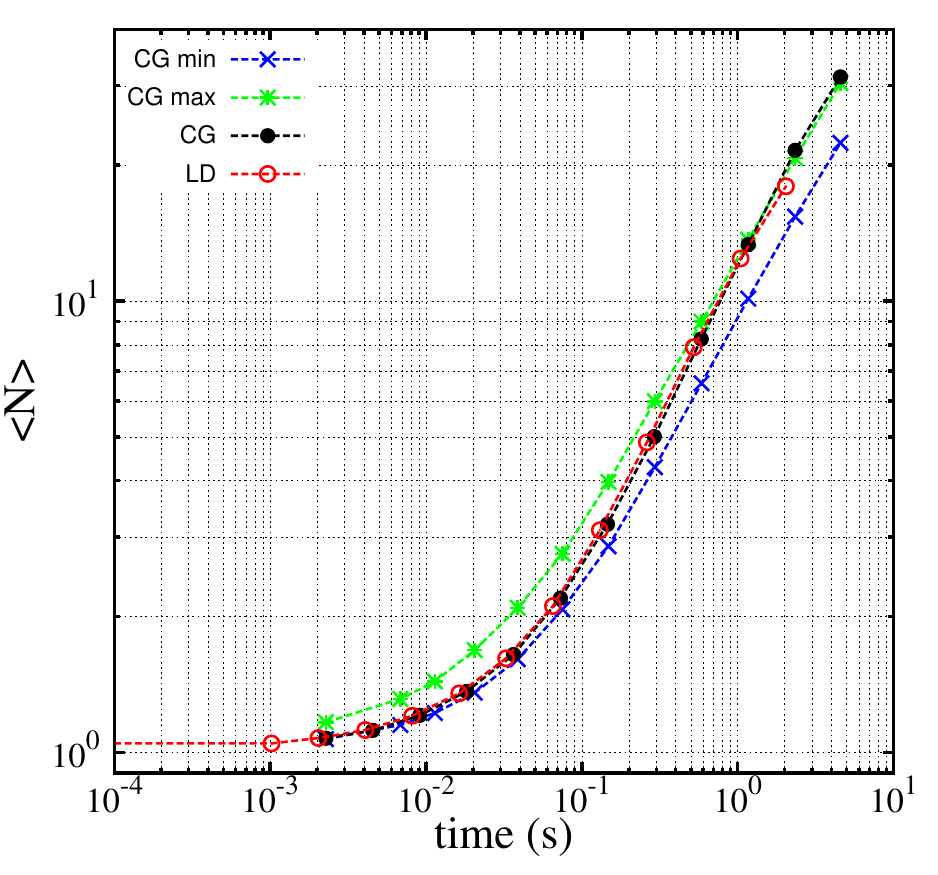}
\caption{(Color online) Comparison of the calculated mean chain size $\langle N\rangle$ between two simplified versions of the coarse-grain model (see main text for details) and the full version. Crosses correspond to the CG40-min simulation, stars to the CG40-max simulation and solid circles correspond to the CG40 (full version) simulation. We also show the Langevin Dynamics results (LD40, open circles).}
\label{fig:comp_attradius}
\end{center}
\end{figure}

All these approaches give us different dynamics of the system as it is shown in Fig.~\ref{fig:comp_attradius} where the mean aggregate size $\langle N\rangle$ is plotted as a function of time together with the corresponding LD results. We observe that the CG40 simulation evolves from an initial behavior close to the CG-min simulation to a behavior closer to the CG-max simulation. Analogous calculations for the CG247 system (not shown here) exhibit identical behavior. In consequence, is important to take into account the full $r_a(s)$ dependence in the simulations as described in our formulation of the model in Section II.

\subsection{An example of practical application: Chain growth and $T_2$ measurements}

\begin{figure}[htp]
\begin{center}
\includegraphics*[width=8cm]{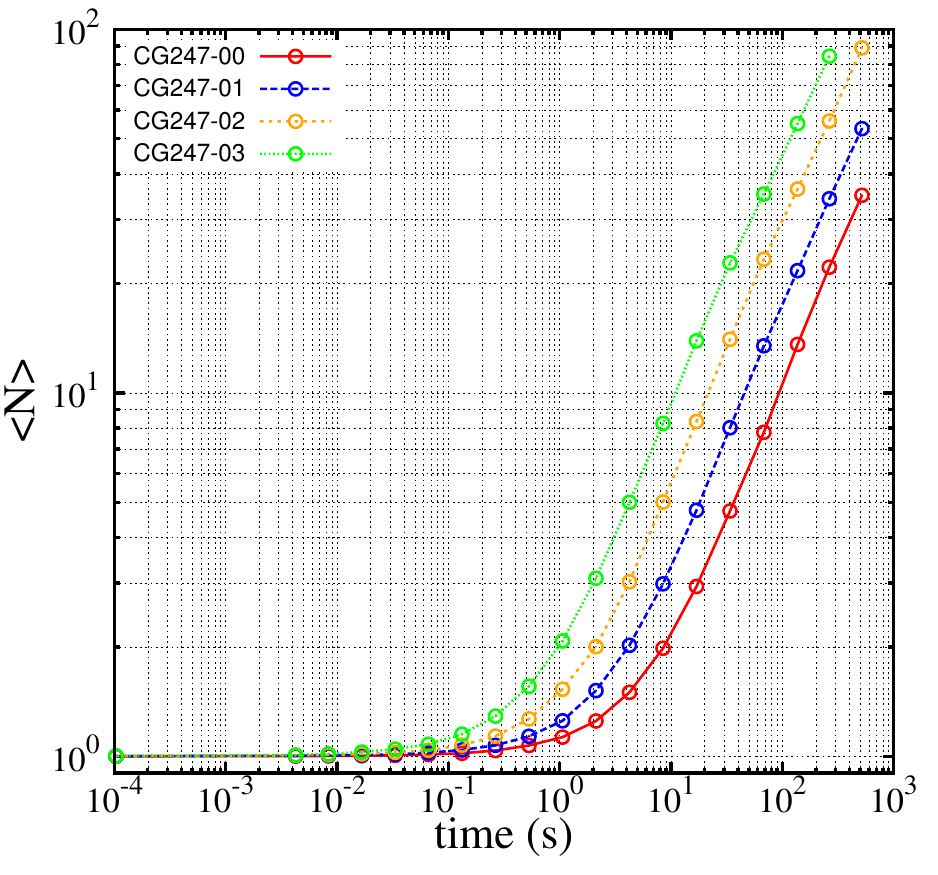}
\caption{(Color online) Average number of colloids in a chain $\langle N\rangle$ as a function of time for the simulations described in Table~\ref{tab:parameters2}.}
\label{fig:comparison_phi}
\end{center}
\end{figure}

\begin{table*}[htdp]
\caption{Set of parameters used for numerical integration in the Coarse Grain simulations of the same colloids described in Case 2 in Table~\ref{tab:colloids} but for different volume fractions $\phi_0$. $N_p$ is the number of particles in the simulation, $L_z$ and $L_x=L_y$ are the sizes of the simulation box (in units of particle diameters) in the directions parallel and perpendicular to the magnetic field, respectively. $\Delta t$ is the time step and $t_f$ is the total simulated time. As in Table~\ref{tab:parameters}, we also indicate the total amount of CPU time employed in the calculation.}
\begin{center}
\begin{tabular}{|c|c|c|c|c|c|c|c|c|c|}
\hline
Label & $N_p$ & $L_z$ & $L_x=L_y$ & $\phi_0$ & $\Delta t$ [s] & $t_f$ [s] & CPU cost \\
\hline
CG247-00 & 8000 & 2436 &1218 & 1.16$\times 10^{-6}$ &1.038$\times 10^{-4}$ & 1221 & 130h1min \\
CG247-01 & 8000 & 1933 &966.5 & 2.32$\times 10^{-6}$ &1.038$\times 10^{-4}$ & 859 & 59h53min \\
CG247-02 & 8000 & 1534 & 767 & 4.64$\times 10^{-6}$ &1.038$\times 10^{-4}$ & 616 & 29h35min \\
CG247-03 & 8000 & 1218 & 609 &  9.28$\times 10^{-6}$ &1.038$\times 10^{-4}$& 806 & 14h52min \\
\hline
\end{tabular}
\end{center}
\label{tab:parameters2}
\end{table*}

Our objective in this subsection is to illustrate the applicability of the methodology developed here in situations of interest for applications of superparamagnetic colloids. As an example, let us consider the use of superparamagnetic colloids as contrasts agents in magnetic resonance imaging. An important issue in this application is the possibility of chain formation of colloids due to the strong magnetic fields applied in the experiments. The formation of chains of colloids in the sample increases the transversal relaxation time $T_2$ of protons, which is an undesired effect in practice. In a previous work, we have employed a preliminary version of our simulation code to analyze this possibility in MRI~\cite{Chen2011}. We have found that under conditions of interest for MRI, significant chaining occurs. We would like to discuss here the results of our simulations as well as compare our results with the experiments in a more direct way than the preliminary simulations presented in Ref.~\cite{Chen2011}.

The system considered here is a dispersion of superparamagnetic colloids in water with the physical properties of Case 2 in Table~\ref{tab:colloids} but now we have considered 4 different values of the initial volume fraction of colloids, according to the experiments in Ref.~\cite{Chen2011}. The corresponding volume fractions are given in Table~\ref{tab:parameters2}. The simulations of these systems, performed with the methodology discussed in section II have been labeled as CG247-00, CG247-01, CG247-02 and CG247-03, respectively and all the technical details are given in Table~\ref{tab:parameters2} (note that the CG247-02 simulation in this table is identical to the simulation CG247 of Table~\ref{tab:parameters}). 

The results for the average number of particles in a chain $\langle N(t)\rangle$ are given in Fig.~\ref{fig:comparison_phi} for the time scales relevant in the experiments of Ref.~\cite{Chen2011}. In all cases we observe significant chain formation even in the case of the smallest concentration. Of course, the kinetics of chain formation is observed to slow down as the concentration of colloids decreases.

The formation of chains has direct impact on the transversal relaxation rate $1/T_2$ of water protons. Initially ($t=0$), the relaxation rate of water protons $1/T^{(0)}_2$ is determined by the presence of a random distribution of isolated (dispersed) colloids. As time goes on, chains form and modify the $T_2$ response of the surrounding water protons. Therefore, the experimentally measured $T_2$ at a given instant $t$ depends on the distribution of chain sizes at that time $t$. As proposed in Ref.~\cite{Chen2011}, we can give a theoretical prediction of the relaxation rate $1/T_2(t)$ from our simulation results by computing the following average:
\begin{equation}
\label{averageR2}
 \frac{1}{T_2(t)}= \frac{1}{N_p}\sum_s s n_s(t) \frac{1}{T^{(s)}_2},
\end{equation}

\begin{figure}[htp]
\begin{center}
\includegraphics*[width=8cm]{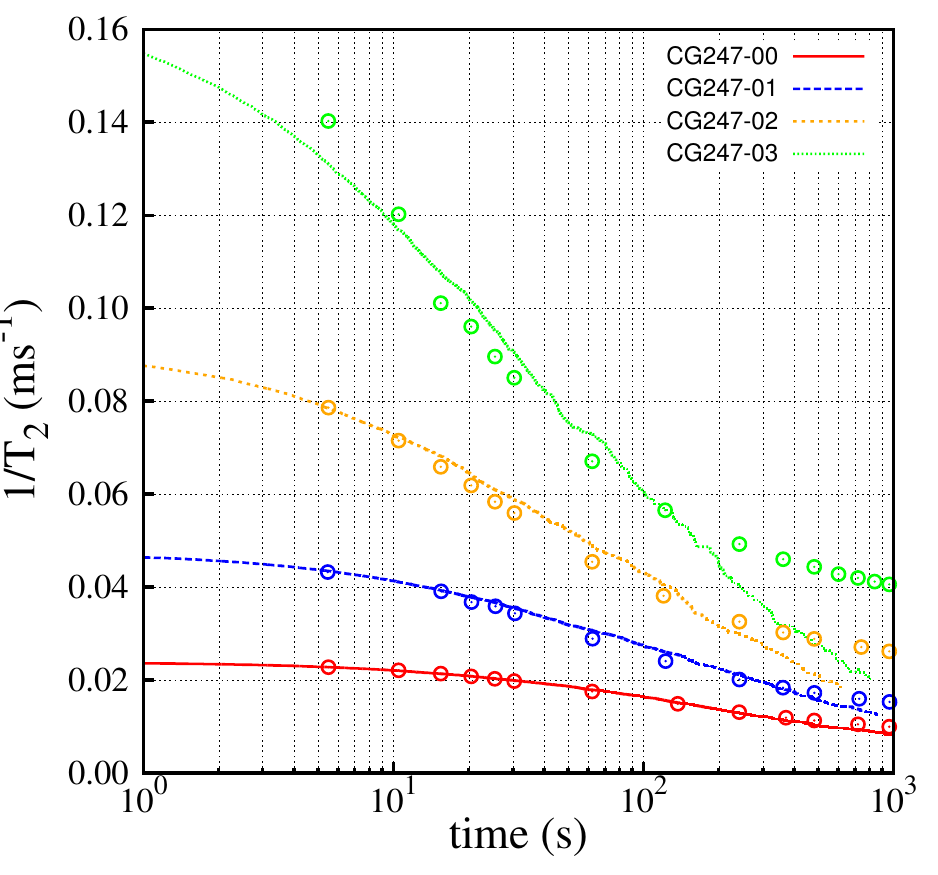}
\caption{(Color online) Evolution of the relaxation rate $1/T_2$ of water protons in 4 dispersions containing different concentrations of superparamagnetic colloids. Solid lines correspond to the predictions obtained from the simulations described in Table~\ref{tab:parameters2} and Eqs.~(\ref{averageR2}) and (\ref{fitR2}). Symbols correspond to experimental data extracted from Fig.5a in Ref~\cite{Chen2011}.}
\label{fig:comparison}
\end{center}
\end{figure}

In Eq.~(\ref{averageR2}), $1/T^{(s)}_2$ is the relaxation rate of water protons near a colloid forming part of a chain containing exactly $s$ colloids and $n_s(t)$ is the number of chains of size $s$ at time $t$, as defined in Section II. Our simulation results provide $n_s(t)$ whereas the calculation of $1/T^{(s)}_2$ requires an additional study of the motion of water protons near a chain containing $s$ colloids. For the particles of the experiments  (Case 2 in our Table~\ref{tab:colloids}), the theoretical results for $1/T^{(s)}_2$ were given in Fig.~9 of Ref.~\cite{Chen2011}. These results can be well fitted to an analytical expression of the form:
 \begin{equation}
\label{fitR2}
 \frac{1}{T_2^{(s)}}=\frac{1}{T_2^{(0)}}s^{-a \cdot s^b}.
\end{equation}
where a fit to the calculations in Ref.~\cite{Chen2011} gives $a=0.0415$ and $b=0.45$. Now, making use of such a fit in Eq.~(\ref{averageR2}) and the values of $n_s(t)$ obtained from simulations CG247-00, CG247-01, CG247-02 and CG247-03, we can make a theoretical prediction for the relaxation rate $1/T_2(t)$. The results are compared in Fig.~\ref{fig:comparison} with experimental results. The simulations show a remarkable agreement between theory and experiments for times corresponding to mean chain length $\langle N\rangle$ larger than 50 colloids. It should be noted that in the case of very long chains, the measurements are not reliable due to sedimentation effects~\cite{Chen2011}.

\section{Conclusions}

In this paper, we have presented a new \emph{on-the-fly} coarse grain model to describe the chaining phenomena observed in dispersions of superparamagnetic colloids under strong external magnetic fields. We report simulation results with the new methodology, which show good agreement with those obtained from more detailed Langevin Dynamics (LD) simulations. The great advantage of the new methodology presented here is its low computational cost in terms of CPU time. As a consequence, we are able to run longer simulations, reaching time scales not accessible in LD simulations. In order to illustrate the applicability of the code in experimentally relevant situations, we have considered the waiting time dependence of the relaxation rate $1/T_2$ of water protons observed in magnetic resonance experiments of dispersions of superparamagnetic colloids \cite{Chen2011}. Experimental results corresponding to waiting times from 1 to 10$^3$ s were correctly predicted by our simulations.

The model, in its present formulation, cannot be applied to situations more complex than irreversible chain growth. However, it seems possible to expand the model to consider other situations of interest. A first generalization could involve the inclusion of lateral interactions between the chains \cite{Furst2000}, which are responsible for the formation of thick chains, observed at volume fractions larger than those considered here \cite{Fermigier1992}. Optical microscopy observations \cite{DeLasCuevas2008} also show the formation of thick chains and bundles in magnetophoresis experiments (motion of magnetic particles under magnetic gradients). Hence, the  inclusion of lateral interactions and deterministic motion of the CG objects will be needed in order to extend our model to study magnetophoresis. Another interesting extension, which is now under way, is the inclusion of the possibility of chain breaking due to thermal fluctuations, a mechanism relevant at low values of the magnetic coupling parameter $\Gamma$. This extension of the model will allow us to study in depth the equilibrium state described in Refs. \cite{Andreu2011,Barrett2011}.

A final improvement to the model could be taking into account the full magnetic response $M(H)$ of the particles in the simulation, in order to simulate situations in which the external magnetic fields are not strong enough to saturate the magnetic colloids. This is a typical situation in many published experimental studies of aggregation of magnetic colloids (see for example~\cite{Promislow1995,Martinez-Pedrero2007}), which focus on the linear magnetic response regime of the colloids.




\section*{Acknowledgments}
This work is supported by the Spanish Government grants FIS2009-13370-C02-02, PET2008-02-81-01, INNPACTO IPT-010000-2010-6 and CONSOLIDER-NANOSELECT-CSD2007-00041. JF and JC are also supported by the Catalan Government grant 2009SGR164. We would like to thank Prof. D.-X. Chen (UAB) for invaluable discussions and help regarding to Eqs.~(\ref{averageR2}) and (\ref{fitR2}).

\appendix
\section{Selection of the integration timestep}

\begin{figure}[htp]
\begin{center}
\includegraphics*[width=8cm]{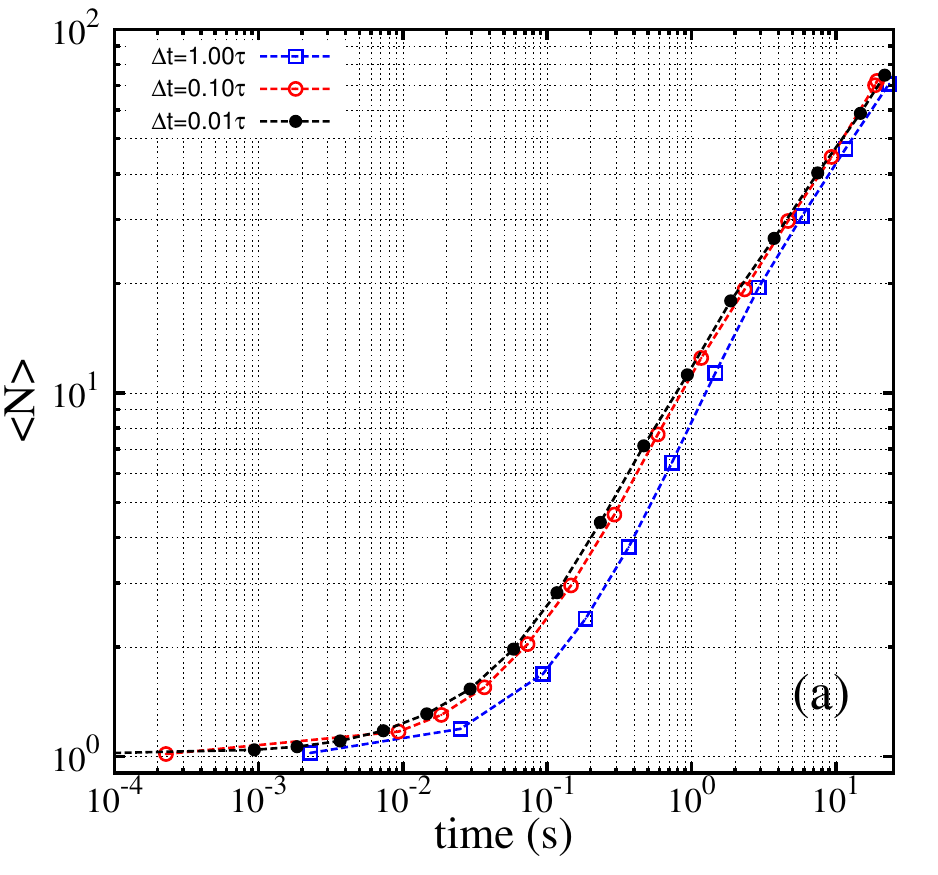}\\
\includegraphics*[width=8cm]{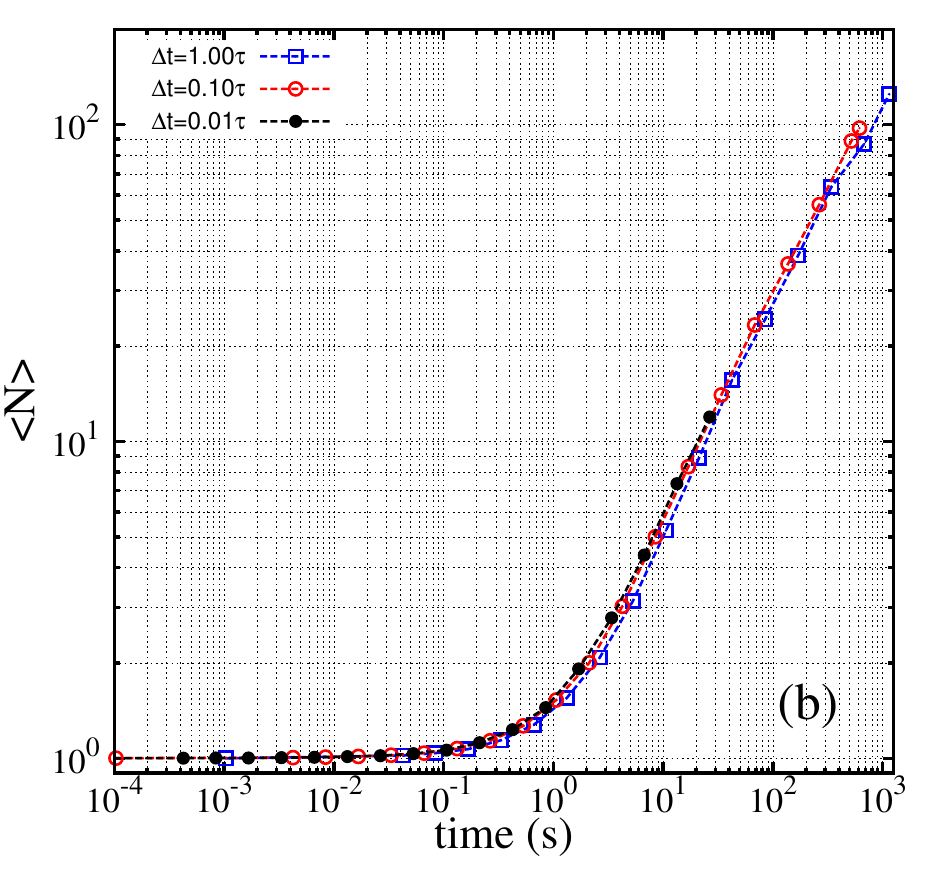}
\caption{(Color online) Influence of the integration timestep $\Delta t$ in the simulations CG40 and CG247 (top and bottom, respectively) on the time evolution of $\langle N\rangle$.}
\label{fig:comp_timestep}
\end{center}
\end{figure}

An important issue that one has to take into account when performing Brownian Dynamics simulations is the proper selection of the integration timestep. The typical diffusive displacement $\ell$ for a single colloid of diameter $d$ and diffusion coefficient $D$ after a timestep $\Delta t$ can be estimated by
\begin{eqnarray}
\label{disp}
\ell \approx \sqrt{6D\Delta t} = d \sqrt{6 \Delta t/ \tau}.
\end{eqnarray}
where we have defined the characteristic diffusion time $\tau \equiv d^2/D$. In general, one selects a timestep $\Delta t$ which results in a displacement $\ell$ smaller than the relevant length scales of the problem (typical separations between particles, range of interaction forces,...).  In our model, the length scale of interactions is given by the radius of the  attraction zones (see Fig.\ref{fig:map}). The typical diffusive displacement $\ell$ corresponding to the selected $\Delta t$ (Eq.(\ref{disp})) has to be smaller than the radius of the attraction zone $r_a$ of the CG objects. In this way, CG objects will correctly explore the attraction zone of other surrounding CG objects. As it is shown in Fig.\ref{fig:map}c, the size $r_a$ of the attraction zone depends on the chain length $s$ (the smallest value of $r_a$ corresponds to $s=1$) and on the coupling parameter $\Gamma$. The dependence on $\Gamma$ is strong, so one has to take into account this fact in selecting $\Delta t$.

In order to check the effect of $\Delta t$ in the results of our simulations at $\Gamma=40$ and $\Gamma=247$, we have repeated the CG40 and CG247 simulations with three different timesteps: $\Delta t$ = 1.00$\tau$, 0.10$\tau$ and 0.01$\tau$. These timesteps $\Delta t$ correspond to typical displacements of $\ell\simeq 2.4d$, $0.8d$ and 0.24$d$ respectively (see Eq.(\ref{disp})). The results of these simulations for the average number of particles in a chain  are shown in Fig.\ref{fig:comp_timestep}. As it can be observed, the effects of the selection of the timestep are critical for the CG40 system and irrelevant for the CG247 system. In order to understand the effect of these different $\Delta t$ , one has to compare the $\ell$ obtained for each $\Delta t$ with the values of $r_a(s=1)$ calculated for $\Gamma=40$ and $\Gamma=247$ (see Figure \ref{fig:map}c). In the CG40 simulation, the smallest radius of a CG attraction zone is $r_a(s=1) = 1.46d$ (see Figure \ref{fig:map}c, case $\Gamma = 40$), so the attraction sphere has a diameter similar to the displacement $\ell = 2.4d$ obtained for $\Delta t$ = 1.0$\tau$. In this case, colloids cannot explore properly the attraction zones and many chain formation events are lost during the simulation run. The other two $\Delta t$ give almost identical results since in both cases $\ell$ is smaller than $r_a(s=1)$, and attraction zones are explored properly. In the CG247 simulations, we observe almost identical results for the three selected $\Delta t$. In this case, we have  $r_a(s=1) = 2.89d$ (see Figure \ref{fig:map}c, case $\Gamma = 247$), which is larger than the corresponding typical displacements $\ell$. As a general rule, if the timestep selected is too large, the chains cannot properly explore the binding sites of the surrounding chains and the chaining process is not correctly simulated.

\end{document}